\documentclass[letterpaper, 10 pt, conference]{ieeeconf}  

\pdfoutput=1

\IEEEoverridecommandlockouts                              

\usepackage{graphics}
\usepackage{epsfig}
\usepackage{times}
\usepackage{amsmath,amssymb,amsfonts}
\allowdisplaybreaks
\usepackage{layout}
\usepackage{algorithm}
\usepackage{algorithmic}
\usepackage{xcolor}
\usepackage{siunitx}
\usepackage{bm}
\usepackage{fp}
\usepackage{tikz}
\usepackage{pgfplots}
\usepackage{eso-pic}

\makeatletter
\let\NAT@parse\undefined
\makeatother
\usepackage{hyperref}

\pgfplotsset{compat=1.18}
\usetikzlibrary{arrows,shapes,backgrounds,patterns,fadings,decorations,
	positioning,external,arrows.meta,pgfplots.fillbetween, pgfplots.groupplots}

\newtheorem{theorem}{Theorem}  
 
\newtheorem{assumption}{Assumption}
\newtheorem{remark}{Remark}

\title{\LARGE \bf
	Learning-Based Optimal Control with Performance Guarantees for Unknown Systems with Latent States}

\author{Robert Lefringhausen$^{1}$, Supitsana Srithasan$^{1}$, Armin Lederer$^{2}$, and Sandra Hirche$^{1}$%
	\thanks{*This work was supported by the European Research Council (ERC) Consolidator Grant "Safe data-driven control for human-centric systems (CO-MAN)" under grant agreement number 864686, by TUM AGENDA 2030, funded by the Federal Ministry of Education and Research (BMBF) and the Free State of Bavaria under the Excellence Strategy of the Federal Government and the Länder as well as by the Hightech Agenda Bavaria, and by the Swiss National Science Foundation under NCCR Automation, grant agreement 51NF40 180545.}%
	\thanks{$^{1}$Chair of Information-oriented Control, TUM School of Computation, Information and Technology, Technical University of Munich, Germany {\tt\small [robert.lefringhausen, supitsana.srithasan, hirche]@tum.de}.}%
	\thanks{$^{2}$Learning and Adaptive Systems Group, Department of Computer Science, ETH Zurich, Switzerland {\tt\small armin.lederer@inf.ethz.ch}.}%
}

\newcommand\DOItext{\footnotesize This is the accepted version of a paper published in the proceedings of the 2024 European Control Conference (ECC). \\ \footnotesize The final published paper can be found at \href{https://doi.org/10.23919/ECC64448.2024.10590972}{doi:10.23919/ECC64448.2024.10590972}.}

\AddToShipoutPicture{\begin{tikzpicture}[remember picture,overlay, align=center]\node[anchor=north,yshift=-15pt] at (current page.north) {\DOItext}; \end{tikzpicture}}

\begin{document}
	
	\maketitle
	\thispagestyle{empty}
	\pagestyle{empty}
	
	\begin{abstract}
		As control engineering methods are applied to increasingly complex systems, data-driven approaches for system identification appear as a promising alternative to physics-based modeling. While the Bayesian approaches prevalent for safety-critical applications usually rely on the availability of state measurements, the states of a complex system are often not directly measurable. It may then be necessary to jointly estimate the dynamics and the latent state, making the quantification of uncertainties and the design of controllers with formal performance guarantees considerably more challenging. This paper proposes a novel method for the computation of an optimal input trajectory for unknown nonlinear systems with latent states based on a combination of particle Markov chain Monte Carlo methods and scenario theory. Probabilistic performance guarantees are derived for the resulting input trajectory, and an approach to validate the performance of arbitrary control laws is presented. The effectiveness of the proposed method is demonstrated in a numerical simulation.
	\end{abstract}
	
	\section{Introduction}
	An accurate mathematical model is fundamental for the model-based control of complex dynamical systems. Since it is often very time-consuming or even impossible to derive physics-based models, e.g., for flexible robotic manipulators or human-robot interaction, data-driven modeling approaches are gaining attention. Accordingly, there are numerous applications in control, with a combination of state-space models (SSMs) and Gaussian processes (GPs) \cite{williams2006} or similar Bayesian approaches being prevalent for safety-critical applications. Their ability to quantify model uncertainties due to limited training data (epistemic uncertainty) and measurement noise (aleatory uncertainty) enables the derivation of probabilistic guarantees for the closed-loop system, even in the presence of noise. Examples can be found, among others, in \cite{umlauft2019} and \cite{beckers2019}, where GP-based tracking controllers are proposed and probabilistic bounds on the tracking error are derived.
	
	A serious disadvantage of the aforementioned methods is that full-state measurements are required. In many applications, however, it is unclear which variables represent the states of the system, or not all of these are measurable. While some learning-based approaches avoid a state-space representation by using past inputs and outputs to make a prediction, e.g., NARX approaches \cite{maiworm2021}, this comes with several drawbacks. First of all, the resulting models cannot separate between noise that permanently changes the trajectory of the system (process noise) and noise that only influences the measurements (measurement noise). They hence provide systematically incorrect open-loop predictions for systems where process and measurement noise differ \cite{schoukens2019}. Another disadvantage is that it is unclear how prior knowledge can be leveraged since it can be expressed almost exclusively in state-space representation. In addition, the states cannot be used for the control task, for example, to formulate constraints. In many cases, it is thus necessary to overcome the dependence of SSMs on state measurements by jointly estimating the dynamics and the latent state. However, to the best of our knowledge, there are no control approaches for nonlinear systems that rigorously account for uncertainties arising from the joint dynamics and state estimation, thus allowing the derivation of performance guarantees in this setting.
	
	The main contribution of this paper is a novel method for the optimal control of unknown systems based on input-output measurements with probabilistic performance guarantees. In order to quantify uncertainties, which is crucial for deriving formal guarantees, we employ a Bayesian approach and formulate a prior over the unknown dynamics and the system trajectory in state-space representation. Since for practical applicability, the prior must be updated based on input-output measurements, but the corresponding posterior distribution is analytically intractable, we utilize particle Markov chain Monte Carlo (PMCMC) methods \cite{andrieu2010} to draw samples from this distribution. While similar approaches are proposed, among others, in \cite{frigola2013}, \cite{tobar2015}, and \cite{svensson2017} for system identification, this fundamental idea has not yet been exploited for control. By employing a scenario point of view \cite{garatti2022}, the resulting samples from the posterior distribution over future system trajectories allow us to analyze the closed-loop performance under an arbitrary fixed control law. Based on this idea, we propose a scenario optimal control problem (OCP), whose solution we prove to exhibit performance and constraint satisfaction guarantees. The straightforward applicability and flexibility of the proposed method are demonstrated through simulations.
	
	The remainder of this paper is structured as follows. After defining the problem in Sec.~\ref{sec:problem_formulation}, Sec.~\ref{sec:state_dynamics_estimation} briefly reviews PMCMC methods. Sec.~\ref{sec:control} describes how the obtained samples can be used to provide formal guarantees for existing control laws and presents a sample-based OCP formulation and guarantees for its solution. The proposed method is numerically evaluated in Sec.~\ref{sec:simulation}, followed by some concluding remarks in Sec.~\ref{sec:conclusion}.
	
	\section{Problem Formulation}
	\label{sec:problem_formulation}
	Consider the general nonlinear discrete-time system of the form\footnote{\textbf{Notation:} Lower/upper case bold symbols denote vectors/matrices, respectively. $\mathbb{R}$ denotes the set of real numbers, $\mathbb{Z}$ the set of integers, $\mathbb{N}^0$ and $\mathbb{N}$ the set of natural numbers with and without zero, and subscripts $_{< a}$ or $_{\leq a}$ the corresponding subsets whose elements satisfy $<a$ or $\leq a$. $\mathbb{P}(\cdot)$ denotes the probability of an event and $p(\cdot)$ the probability density function (pdf). (Multivariate) Gaussian distributed random variables with mean $\bm{\mu}$ and variance $\bm{\Sigma}$ are denoted by $\mathcal{N}(\bm{\mu},\bm{\Sigma})$. $\bm{a}_{b \in \mathbb{Z}:c \in \mathbb{Z}}$ is the shorthand notation for $\{ \bm{a}_{b}, \dots, \bm{a}_{c}\}$. $\bm{I}_n$ denotes the $n \times n$ identity matrix and $\bm{0}$ the zero matrix of appropriate dimension. $|\cdot|$ denotes the absolute value of a scalar and $\operatorname{card}(\cdot)$ the cardinality of a set.}
	\vspace{-0.4cm}
	\begin{subequations} \label{eq:system}
		\begin{align} 
			\bm{x}_{t+1} &= \bm{f}(\bm{x}_t,\bm{u}_t)+\bm{v}_t,\\
			\bm{y}_{t} &= \bm{g}(\bm{x}_t,\bm{u}_t)+\bm{w}_t,
		\end{align}
	\end{subequations}
	with state $\bm{x}_t \in \mathbb{R}^{n_x\in \mathbb{N}}$, which is not observed explicitly, input $\bm{u}_t \in \mathbb{R}^{n_u\in \mathbb{N}}$, output $\bm{y}_t \in \mathbb{R}^{n_y\in \mathbb{N}}$, and $t \in \mathbb{Z}$. The system is perturbed by independent and identically distributed (iid) process noise $\bm{v}_t \sim \mathcal{V}$ and measurement noise $\bm{w}_t \sim \mathcal{W}$, which are independent of the state and the input. The transition and observation functions $\bm{f}(\cdot)$ and $\bm{g}(\cdot)$ as well as the noise distributions $\mathcal{V}$ and $\mathcal{W}$ are assumed to be unknown. We assume that at time $t=0$ we have access to the dataset $\mathbb{D}=\{\bm{u}_{t}, \bm{y}_{t}\}_{t=T\in \mathbb{Z}_{<-1}:-1}$ containing the last $|T|$ input-output measurements. In order to enable the theoretically justified inference of a model and to provide formal guarantees for its generalization, we further employ the following assumption.
	
	\begin{assumption}
		\label{as:sample_from_prior}
		The structure of the system (\ref{eq:system}) is known, i.e., suitable parameterizations $\{ \bm{f}_{\bm{\theta}}(\cdot),\bm{g}_{\bm{\theta}}(\cdot),\allowbreak \mathcal{V}_{\bm{\theta}},\allowbreak \mathcal{W}_{\bm{\theta}}\}$ with a finite number of unknown parameters $\bm{\theta}$ are available. In addition, priors $p(\bm{\theta})$ and $p(\bm{x}_{T})$ for the model parameters and the initial state of the observed trajectory are available\footnote{Note that the prior for the model parameters $\bm{\theta}$ and the initial state $\bm{x}_{T}$ also implies a prior for all subsequent states.}. The unknown system and the initial state are samples from these priors.
	\end{assumption}
	
	Knowledge of a reasonable prior is a standard assumption in Bayesian inference. In many cases, models for $\bm{f}(\cdot)$ and $\bm{g}(\cdot)$ can be constructed based on physical principles, e.g., for cooperative manipulation \cite{dohmann2021}. Moreover, if such knowledge is unavailable, methods from Gaussian process regression can be used to construct parametric models, which can asymptotically represent all continuous functions with an increasing number of basis functions, such as the approach presented in \cite{solin2020}. Often, the type of noise is also known, and only the parameters of the corresponding distribution must be determined. Therefore, this assumption is not particularly restrictive in practice.
	
	Given a stage cost $c(\bm{u}_t, \bm{x}_t, \bm{y}_t)$, the objective is to find an input trajectory $\bm{u}_{0:H}$ that minimizes the accumulated cost over the horizon $H \in \mathbb{N}$
	\begin{equation} \label{eq:cost}
		J_H = \sum_{t=0}^{H}c\left(\bm{u}_t, \bm{x}_t, \bm{y}_t\right)
	\end{equation}
	while satisfying constraints of the form
	\begin{equation} \label{eq:constraints}
		h\left(\bm{u}_{0:H},\bm{x}_{0:H},\bm{y}_{0:H}\right)\leq 0,
	\end{equation}
	where $h(\cdot)$ is an arbitrary deterministic function. Since the unknown parameters and the process and measurement noise may have infinite support, it is generally impossible to guarantee the satisfaction of the constraints (\ref{eq:constraints}) in every case, i.e., with probability $1$. Moreover, input trajectories with such guarantees tend to be overly conservative and are thus undesirable due to poor performance. Instead, we aim to find an input trajectory that is probabilistically robust, i.e., for which we can derive guarantees of the form $\mathbb{P}(h(\bm{u}_{0:H},\bm{x}_{0:H},\bm{y}_{0:H})\leq 0) \geq \alpha$ when applied to the system. In order to address this problem, we first consider how probabilistic guarantees can be provided for the case that a given control law $\bm{u}_t = \bm{\pi}(\bm{u}_{T:t-1},\bm{y}_{T:t-1},t)$ is applied to the unknown system. Based on this analysis, we consider how the search for an input trajectory $\bm{u}_{0:H}$ can be formulated as a tractable OCP for whose solution probabilistic performance and constraint satisfaction guarantees can be inferred directly.
	
	\section{Particle Markov chain Monte Carlo methods}
	\label{sec:state_dynamics_estimation}
	For practical applicability, the prior presented in the previous section must be updated based on the observations $\mathbb{D}$, i.e., the posterior distribution over model parameters and latent state trajectories\footnote{Note that throughout the work, the dependency on the prior is omitted for notational simplicity.} $p(\bm{\theta}, \bm{x}_{T:-1} \mid \mathbb{D})$ must be inferred. Without inferring the posterior distribution, the stochastic OCP is usually infeasible since the prior over system trajectories, which implicitly follows from the repeated propagation of $p(\bm{x}_{T})$ through the prior distribution over the dynamics, exhibits an excessive variance. In general, inference of the posterior distribution $p(\bm{\theta}, \bm{x}_{T:-1} \mid \mathbb{D})$ is analytically intractable, but particle Markov chain Monte Carlo (PMCMC) methods, introduced in \cite{andrieu2010} and briefly reviewed in this section, can be used to draw samples from it.
	
	PMCMC methods utilize sequential Monte Carlo (SMC) algorithms \cite{doucet2009}, which numerically approximate the distribution over the latent state trajectory $p(\bm{x}_{T:-1} \mid \bm{\theta}, \mathbb{D})$ by a set of $N$ weighted particles, to generate proposals for Markov chain Monte Carlo (MCMC) samplers, a broad class of algorithms designed for sampling from posterior distributions.
	
	An example is particle Gibbs (PG) sampling, which enables particularly efficient sampling from $p(\bm{\theta}, \bm{x}_{T:-1} \mid \mathbb{D})$ in case samples can be drawn from the conditional distribution $p(\bm{\theta} \mid \bm{x}_{T:-1}, \mathbb{D})$. In PG sampling, a state trajectory $\bm{x}_{T:-1}^{[k]}$ is first drawn from the distribution $p(\bm{x}_{T:-1}^{[k]} \mid \bm{\theta}^{[k]}, \mathbb{D})$ using an SMC algorithm. Based on this state trajectory, new model parameters $\bm{\theta}^{[k+1]}$ are then drawn from $p(\bm{\theta}^{[k+1]} \mid \bm{x}_{T:-1}^{[k]}, \mathbb{D})$. These two steps are repeated until the desired number of samples is reached.
	
	\begin{remark}
		Note that the distribution $p(\bm{\theta} \mid \bm{x}_{T:-1}, \mathbb{D})$, required for PG sampling, depends on the prior $p(\bm{\theta})$. Various combinations of parametric models and priors can be found in the literature that provide a closed-form expression for $p(\bm{\theta} \mid \bm{x}_{T:-1}, \mathbb{D})$ and thus enable efficient PG sampling.
	\end{remark}
	
	\begin{remark}
		If the conditional distribution $p(\bm{\theta} \mid \bm{x}_{T:-1}, \mathbb{D})$ is not available, other PMCMC methods, e.g., particle marginal Metropolis-Hastings sampling \cite{andrieu2010}, can be applied. The theory presented hereafter is not tied to a specific PMCMC method.
	\end{remark}
	
	It can be shown that the invariant distribution of PG sampling and other suitable PMCMC methods is $p(\bm{\theta}, \bm{x}_{T:-1} \mid \mathbb{D})$ and that they will asymptotically provide samples from this distribution even for a finite number of particles \cite{andrieu2010}. Since the first samples depend heavily on the initialization and may not accurately represent the desired distribution, the first $K_b$ samples must be discarded (so-called burn-in period). As for all MCMC methods, no guarantees can be given regarding the length of the burn-in period. The samples are also correlated with samples generated immediately before and after. Since independent samples are desired for the intended application, additional measures must be taken to reduce the correlation. Several approaches are presented in the literature; see, e.g., \cite{tierney1994}. A straightforward and comparatively sample-efficient approach is thinning, which reduces the correlation by using only every $k_d$-th sample. The parameter $k_d$ should be chosen as small as possible so that few samples are discarded but large enough so that the samples are approximately independent. An indicator of an appropriate $k_d$ is a low auto-correlation of successive samples.
	
	\begin{remark}
		How well a PMCMC algorithm can explore the posterior distribution and how small $k_d$ can be chosen also depends significantly on the prior and the parameters of the algorithm. In order to reduce $k_d$ to an acceptable level, scaling the model parameters or adjusting the prior may be necessary.
	\end{remark}
	
	With a burn-in period of appropriate length and a suitable measure to reduce the correlation, PMCMC methods generate almost uncorrelated samples from the desired distribution, justifying the following assumption.
	
	\begin{assumption}
		\label{as:iid_posterior}
		The employed PMCMC method provides $K$ independent samples $\{ \bm{\theta},\bm{x}_{T:-1}\} ^{[1:K]}$ from the distribution $p(\bm{\theta}, \bm{x}_{T:-1} \mid \mathbb{D})$.
	\end{assumption}

	\section{Control with Guarantees}
	\label{sec:control}
	In this section, we consider how input trajectories with probabilistic performance guarantees can be derived based on the prior over the system dynamics and the latent state trajectory presented in Sec.~\ref{sec:problem_formulation}. This prior is very general and can accurately represent a wide range of systems due to the separation of process and measurement noise. In addition, the state-space representation makes it very intuitive to incorporate prior knowledge. In order to determine an input trajectory, we propose to utilize PMCMC methods to sample from the posterior distribution over future trajectories of the unknown system, which depends on the control inputs $\bm{u}_{0:H}$. Due to the representation as a parametric state-space model, future state and output trajectories can be expressed as a function of $\bm{u}_{0:H}$, which is uniquely characterized by the model parameters $\bm{\theta}$, the initial state $\bm{x}_0$, and the realizations of the process and measurement noise $\bm{v}_{0:H}$ and $\bm{w}_{0:H}$, respectively. Samples $\bm{\delta}$ from the corresponding posterior distribution $p(\bm{\theta}, \bm{x}_0, \bm{v}_{0:H}, \bm{w}_{0:H}\mid \mathbb{D})$ represent possible future system behavior depending on $\bm{u}_{0:H}$ and we refer to these samples as scenarios in the following. Algorithm~\ref{alg:scenario_generation} generates the scenarios based on the observations $\mathbb{D}$. First, $K$ model parameters $\bm{\theta}$ and state trajectories $\bm{x}_{T:-1}$ are drawn from the corresponding posterior distribution using a PMCMC method (line 2). Then, possible realizations of the process and measurement noise are drawn from the distributions parameterized by $\bm{\theta}$ (lines 3-6), and the initial state is determined via forward propagation (line 7).
	
	The scenarios obtained by Algorithm~\ref{alg:scenario_generation} depict not only the aleatory uncertainty but also the epistemic uncertainty. As we will explain in the following, these scenarios are a powerful tool for analyzing and synthesizing control laws. In Sec.~\ref{sub:control_analysis}, we first outline how they can be used to provide probabilistic guarantees for existing control laws. Then, in Sec.~\ref{sub:optimal_control}, we present a scenario-based OCP formulation and derive probabilistic guarantees for its solution.

	\begin{algorithm}[t]
		\caption{Scenario generation}\label{alg:scenario_generation}
		\begin{algorithmic}[1]
			\renewcommand{\algorithmicrequire}{\textbf{Input:}}
			\renewcommand{\algorithmicensure}{\textbf{Output:}}
			\REQUIRE Dataset $\mathbb{D}$, parametric model $\{\bm{f}_{\bm{\theta}}(\cdot),\bm{g}_{\bm{\theta}}(\cdot), \mathcal{V}_{\bm{\theta}}, \mathcal{W}_{\bm{\theta}}\}$, priors $p(\bm{\theta})$ and $p(\bm{x}_{T})$, $K$, $H$ 
			\ENSURE Scenarios $\bm{\delta}^{[1:K]} = \{\bm{\theta},\bm{x}_{0},\bm{v}_{0:H},\bm{w}_{0:H}\}^{[1:K]}$
			\FOR {$k = 1$ to $K$}
			\STATE Sample $\{\bm{\theta}, \bm{x}_{T:-1}\}^{[k]}$ from $p(\bm{\theta}, \bm{x}_{T:-1} \mid \mathbb{D})$ using\\ a PMCMC method.
			\FOR {$t = -1$ to $H$}
			\STATE Sample $\bm{v}_t^{[k]}$ from $ \mathcal{V}_{\bm{\theta}^{[k]}}$.
			\STATE Sample $\bm{w}_t^{[k]}$ from $ \mathcal{W}_{\bm{\theta}^{[k]}}$.	
			\ENDFOR
			\STATE Set $\bm{x}_0^{[k]} = \bm{f}_{\bm{\theta}^{[k]}}(\bm{x}_{-1}^{[k]}, \bm{u}_{-1} ) +\bm{v}_{-1}^{[k]}$.
			\ENDFOR
		\end{algorithmic}
	\end{algorithm}
	
	\subsection{Analysis of existing control laws}
	\label{sub:control_analysis}
	In the following, we show how the scenarios $\bm{\delta}^{[1:K]}$ can be used to provide probabilistic performance and constraint satisfaction guarantees for arbitrary control laws $\bm{u}_t = \bm{\pi}(\bm{u}_{T:t-1},\bm{y}_{T:t-1},t)$. We thereby use a very natural idea to analyze the closed-loop behavior: since the scenarios represent samples from the distribution over possible future system trajectories depending on the inputs, we can obtain samples $\{\bm{u}_{0:H}, \bm{x}_{0:H}, \bm{y}_{0:H}, J_H\}^{[1:K]}$ from the posterior distribution over future inputs, states, outputs, and costs by simulating the scenarios together with the control law forward as outlined in Algorithm~\ref{alg:forward_simulation}. In order to enable a formal performance analysis based on these samples, the following prerequisite must be fulfilled.
	
	\begin{algorithm}[t]
		\caption{Forward simulation}\label{alg:forward_simulation}
		\begin{algorithmic}[1]
			\renewcommand{\algorithmicrequire}{\textbf{Input:}}
			\renewcommand{\algorithmicensure}{\textbf{Output:}}
			\REQUIRE Scenarios $\bm{\delta}^{[1:K]}$, control law $\bm{\pi}(\bm{u}_{T:t-1},\bm{y}_{T:t-1},t )$, $K$, $H$
			\ENSURE Samples $\{\bm{u}_{0:H}, \bm{x}_{0:H}, \bm{y}_{0:H}, J_H \}^{[1:K]}$
			\FOR {$k = 1$ to $K$}
			\FOR {$t = 0$ to $H$}
			\STATE Set $\bm{u}_t^{[k]}=\bm{\pi}(\bm{u}_{T:t-1}^{[k]},\bm{y}_{T:t-1}^{[k]},t)$.
			\STATE Set $\bm{x}_{t+1}^{[k]} = \bm{f}_{\bm{\theta}^{[k]}}(\bm{x}_t^{[k]} ,\bm{u}_t^{[k]})+\bm{v}_t^{[k]}$.
			\STATE Set $\bm{y}_{t}^{[k]} = \bm{g}_{\bm{\theta}^{[k]}}( \bm{x}_{t}^{[k]} ,\bm{u}_{t}^{[k]})+\bm{w}_t^{[k]}$.
			\ENDFOR
			\STATE Set $J_H^{[k]}=J_H( \bm{u}_{0:H}^{[k]}, \bm{x}_{0:H}^{[k]}, \bm{y}_{0:H}^{[k]})$.
			\ENDFOR
		\end{algorithmic}
	\end{algorithm}
	
	\begin{assumption}
		\label{as:indepentent_control law}
		The control law $\bm{u}_t = \bm{\pi}(\cdot)$ is independent of the scenarios $\bm{\delta}^{[1:K]}$, which are used to analyze it, i.e., it is not designed based on these scenarios.
	\end{assumption}
	
	This assumption is not limiting since additional scenarios can be generated for the analysis if scenario-dependent control laws are to be investigated.
	
	In order to derive probabilistic guarantees for the cost incurred when the control law is applied to the unknown system, we consider the maximum of the $K$ samples of the accumulated cost (\ref{eq:cost}) generated by Algorithm~\ref{alg:forward_simulation}
	\begin{equation}\label{eq:worst_case_cost}
		\overline{J_H} = \max_{k \in \mathbb{N}_{\leq K}} J_H^{[k]}.
	\end{equation}
	The following theorem provides a statement on the probability that this worst-case cost $\overline{J_H}$, which is a random variable itself, is an upper bound for the incurred cost.
	
	\begin{theorem}
		\label{theo:cost_bound}
		For a given confidence parameter  $\beta \in (0,1)$, under Assumptions \ref{as:sample_from_prior}, \ref{as:iid_posterior}, and \ref{as:indepentent_control law}, it holds that
		\begin{equation}
			\mathbb{P}\left(V_J \leq 1-\sqrt[K]{\beta}\right)\geq 1-\beta,
		\end{equation}
		where $V_J$ denotes the probability that the cost (\ref{eq:cost}), incurred when the control law $\bm{\pi}(\cdot)$ is applied to the unknown system, exceeds $\overline{J_H}$, i.e., $ V_J = \mathbb{P}(J_H>\overline{J_H})$.
		
		\proof
		Due to Assumptions \ref{as:sample_from_prior} and \ref{as:iid_posterior}, the samples $\{\bm{\theta},\bm{x}_{T:-1}\}^{[1:K]}$ generated by the PMCMC method are independent samples from the posterior distribution over system dynamics and past state trajectories. Due to Assumption~\ref{as:indepentent_control law}, the samples $\{\bm{u}_{0: H}, \bm{x}_{0:H}, \bm{y}_{0:H}, J_H\}^{[1:K]}$ generated with Algorithms~\ref{alg:scenario_generation} and \ref{alg:forward_simulation} are therefore also independent samples from the respective posterior distributions. $\overline{J_H}$ corresponds to the solution of a sample-dependent optimization problem with the single decision variable $\overline{J_H}$, the objective $\min \overline{J_H}$, which is convex in the decision variable, and the constraint $\max_{k \in \mathbb{N}_{\leq K}} J_H^{[k]}-\overline{J_H} \leq0$, which is also convex in $\overline{J_H}$. The result follows from \cite[Theorem 1]{campi2008} by simple algebraic manipulations.
		\endproof
	\end{theorem}
	
	If the constraints (\ref{eq:constraints}) are satisfied for all scenarios, the same line of thought can be used to derive guarantees that the constraints are satisfied when the control law $\bm{\pi}(\cdot)$ is applied to the unknown system.
	
	\begin{theorem}
		\label{theo:constraints_convex}
		Assume that with a control law $\bm{\pi}(\cdot)$ the constraints (\ref{eq:constraints}) are satisfied for all scenarios, i.e.,
		\begin{equation}
			h\left(\bm{u}_{0:H}^{[k]},\bm{x}_{0:H}^{[k]},\bm{y}_{0:H}^{[k]}\right)\leq 0 \quad \forall k \in \mathbb{N}_{\leq K}
		\end{equation}
		where $\{ \bm{u}_{0:H},\bm{x}_{0:H},\bm{y}_{0:H} \} ^{[1:K]}$ are obtained via Algorithm~\ref{alg:forward_simulation}.
		For a given confidence parameter  $\beta \in (0,1)$, under Assumptions \ref{as:sample_from_prior}, \ref{as:iid_posterior}, and \ref{as:indepentent_control law}, it holds that
		\begin{equation}
			\mathbb{P}\left(V_h \leq 1-\sqrt[K]{\beta}\right)\geq 1-\beta,
		\end{equation}
		where $V_h$ denotes the probability that the constraints (\ref{eq:constraints}) are violated when the control law $\bm{\pi}(\cdot)$ is applied to the unknown system.
		
		\proof
		This follows from the proof of Theorem~\ref{theo:cost_bound}, if one considers $\max_{k \in \mathbb{N}_{\leq K}} h(\bm{u}_{0:H}^{[k]},\bm{x}_{0:H}^{[k]},\bm{y}_{0:H}^{[k]})\leq 0$ instead of the maximum cost $\overline{J_H}$.
		\endproof
	\end{theorem}
	
	Theorems \ref{theo:cost_bound} and \ref{theo:constraints_convex} thus provide probabilistic performance and constraint satisfaction guarantees when arbitrary control laws are applied to the unknown system. The strength of the guarantees depends on the total number of samples $K$, and more samples lead to stronger guarantees.
	
	\subsection{Scenario-based optimal control}
	\label{sub:optimal_control}
	In the previous section, we have shown that a finite number of scenarios $\bm{\delta}^{[1:K]}$ can be used to provide probabilistic guarantees for existing control laws. A natural question is whether these scenarios can also be used directly to find a control law or input trajectory that robustly minimizes the cost (\ref{eq:cost}) while satisfying the constraints (\ref{eq:constraints}) with high probability. As we will show in the following, it is possible to formulate the search for such an input trajectory\footnote{Note that the considerations given here can be readily extended to finding parameters of arbitrary parametric control laws $\bm{\pi}(\cdot)$.} $\bm{u}_{0:H}$ as a deterministic OCP for whose solution probabilistic constraint satisfaction guarantees can be directly provided. For this purpose, we make use of the ideas presented in Sec.~\ref{sub:control_analysis} and formulate the following OCP:
	\begin{subequations} \label{eq:scenario_OCP}
		\begin{align} 
			&\min_{\bm{u}_{0:H},\; \overline{J_H}} \overline{J_H} \label{eq:scenario_OCP_cost}\\
			\text{sub}&\text{ject to:} \quad \forall k  \in \mathbb{N}_{\leq K}, \quad \forall t  \in \mathbb{N}^{0}_{\leq H} \nonumber\\
			& \bm{x}_{t+1}^{[k]} = \bm{f}_{\bm{\theta}^{[k]}}\left(\bm{x}_t^{[k]} ,\bm{u}_t\right)+\bm{v}_t^{[k]}, \label{eq:scenario_OCP_dynamic_constraints_1}\\
			&\bm{y}_{t}^{[k]} = \bm{g}_{\bm{\theta}^{[k]}}\left( \bm{x}_{t}^{[k]} ,\bm{u}_{t}\right)+\bm{w}_t^{[k]}, \label{eq:scenario_OCP_dynamic_constraints_2} \\
			& J_H^{[k]}=J_H\left( \bm{u}_{0:H}, \bm{x}_{0:H}^{[k]}, \bm{y}_{0:H}^{[k]}\right) \leq \overline{J_H}, \label{eq:scenario_OCP_cost_constraints}\\
			&h\left(\bm{u}_{0:H},\bm{x}_{0:H}^{[k]},\bm{y}_{0:H}^{[k]}\right) \leq 0. \label{eq:scenario_OCP_constraints}
		\end{align}
	\end{subequations}
	In order to ensure the robust minimization of the cost (\ref{eq:cost}), the objective of the OCP is to minimize the worst-case cost $\overline{J_H}$, i.e., similar to (\ref{eq:worst_case_cost}), the costs of all scenarios must be less than or equal to $\overline{J_H}$ resulting in (\ref{eq:scenario_OCP_cost_constraints}). The forward simulation utilized in Sec.~\ref{sub:control_analysis} (Algorithm~\ref{alg:forward_simulation}) is implicitly included in the optimization problem via the constraints (\ref{eq:scenario_OCP_dynamic_constraints_1}-\ref{eq:scenario_OCP_dynamic_constraints_2}). Similar to the previous section, we also require that the constraints (\ref{eq:constraints}) are satisfied for all scenarios, resulting in (\ref{eq:scenario_OCP_constraints}).
	
	The optimization problem (\ref{eq:scenario_OCP}) is deterministic, more precisely, a nonlinear programming (NLP) problem in which the uncertainty about the unknown system is incorporated by considering multiple deterministic scenarios rather than in terms of distributions. The problem can thus be solved with well-known methods; see \cite{nocedal2006} for an overview. Since the problem (\ref{eq:scenario_OCP}) is generally non-convex, convergence to the global minimum cannot be guaranteed. However, locally optimal solutions are also well-suited for many practical applications.
	
	In the following, we assume that a feasible and locally optimal solution $\{\bm{u}_{0:H}^\star, \overline{J_H}^\star\}$ to the problem (\ref{eq:scenario_OCP}) is available. Note that epistemic and aleatory uncertainties play a crucial role here. In (\ref{eq:scenario_OCP_constraints}), it is required that the constraints (\ref{eq:constraints}) are satisfied for all scenarios. If the sampled scenarios are very different from each other and the uncertainty is thus high, finding a feasible input trajectory may be challenging. However, it is also generally expected that if the uncertainty about the system's behavior is high, it may be difficult or impossible to find an input that satisfies the constraints with high probability - the proposed method is no exception. 
	\begin{remark}
		In general, the uncertainty about the latent states $\bm{x}_{0:H}$ is much higher than about the observable output $\bm{y}_{0:H}$ or $\bm{g}(\bm{x}_{0:H},\bm{u}_{0:H})$. Therefore, in case of high uncertainty (e.g., many unknown parameters and a flat prior), the constraints (\ref{eq:constraints}) should depend only on $\bm{u}_{0:H}$, $\bm{y}_{0:H}$, and $\bm{g}(\bm{x}_{0:H},\bm{u}_{0:H})$. In practice, the constraints will usually depend only on these variables since, without measurability, it is difficult to formulate meaningful constraints for the states.
	\end{remark}
	
	Similar to Sec.~\ref{sub:control_analysis}, we are interested in deriving probabilistic guarantees for the incurred cost and the satisfaction of constraints when the input trajectory $\bm{u}_{0:H}^\star$, obtained by solving the OCP (\ref{eq:scenario_OCP}), is applied to the unknown system. In (\ref{eq:scenario_OCP}), only a finite number of possible future system trajectories are considered, rather than the corresponding analytically intractable distribution. However, in a safety-critical context, ensuring that the constraints (\ref{eq:constraints}) are satisfied for all possible trajectories with high probability is necessary. Theorems \ref{theo:cost_bound} and \ref{theo:constraints_convex} do not apply in this case since the control input depends on the scenarios, and therefore Assumption~\ref{as:indepentent_control law} is not fulfilled. In order to obtain a formal statement about the worst-case cost and the probability of a constraint violation, we instead consider variants of the optimization problem (\ref{eq:scenario_OCP}) where the constraints (\ref{eq:scenario_OCP_dynamic_constraints_1}-\ref{eq:scenario_OCP_constraints}) are in effect for only a subset of the scenarios. In order to derive formal guarantees by solving these problems, we need to ensure that the solver used to solve the OCP (\ref{eq:scenario_OCP}) provides repeatable answers. This is formalized in the following assumption \cite{garatti2021}.
	
	\begin{assumption}
		\label{as:solver}
		The OCP (\ref{eq:scenario_OCP}) and the employed solver represent a mapping $M_{K}$ from the scenarios $\bm{\delta}^{[1:K]}$ to a solution $\{\bm{u}_{0:H}^\star, \overline{J_H}^\star\}$. Let $\mathbb{H}_{\bm{\delta}^{[i]}}$ denote the set that contains all solutions that are suitable for a scenario $\bm{\delta}^{[i]}$, i.e., all $\{\bm{u}_{0:H}, \overline{J_H}\}$ that satisfy the constraints (\ref{eq:scenario_OCP_dynamic_constraints_1}-\ref{eq:scenario_OCP_constraints}) for $k=i$. Then, for every $K\in \mathbb{N}$, every $k\in \mathbb{N}$, and for every choice of $\bm{\delta}^{[1]},\dots,\bm{\delta}^{[K]}$ and $\bm{\delta}^{[K+1]},\dots,\bm{\delta}^{[K+k]}$, the following three properties hold: 
		\begin{enumerate}
			\item[(i)] If $\bm{\delta}^{[i_1]},\dots, \bm{\delta}^{[i_{K}]}$ is a permutation of $\bm{\delta}^{[1]},\dots,\bm{\delta}^{[{K}]}$, then $ M_{K}(\bm{\delta}^{[1]},\dots,\bm{\delta}^{[{K}]}) = M_{K}(\bm{\delta}^{[i_1]},\dots, \bm{\delta}^{[i_{K}]})$.
			\item[(ii)] If $M_{K}(\bm{\delta}^{[1]},\dots,\bm{\delta}^{[{K}]}) \in \mathbb{H}_{\bm{\delta}^{[K+i]}}$ for all $ i = 1,\dots,k$, then $ M_{K}(\bm{\delta}^{[1]},\dots,\bm{\delta}^{[{K}]}) = M_{K+k}(\bm{\delta}^{[1]},\dots,\bm{\delta}^{[K+k]})$.
			\item[(iii)] If $ M_{K}(\bm{\delta}^{[1]},\dots,\bm{\delta}^{[{K}]}) \notin \mathbb{H}_{\bm{\delta}^{[K+i]}}$ for at least one $i =1,\dots,k$, then $ M_{K}(\bm{\delta}^{[1]},\dots,\bm{\delta}^{[{K}]}) \neq M_{K+k}(\bm{\delta}^{[1]},\dots,\bm{\delta}^{[K+k]})$.
		\end{enumerate}
	\end{assumption}
	
	This assumption is not restrictive, and it can be straightforwardly shown to be satisfied for most deterministic gradient-based solvers, provided that the same initialization is used and the cost function has a unique local minimum, possibly after inserting a regularization term. However, it is generally not fulfilled for random-based methods such as genetic algorithms, which may provide different solutions when considering the same scenarios due to their reliance on randomness.
	
	Now suppose a set $\mathbb{S} \subset \mathbb{N}_{\leq K}$ with cardinality $s:= \operatorname{card}(\mathbb{S}) < K$ is known for which a solver satisfying Assumption~\ref{as:solver} yields the same solution as for the original OCP (\ref{eq:scenario_OCP}) when only the scenarios belonging to this set are considered. That is, the solution of the OCP (\ref{eq:scenario_OCP}), where the constraints (\ref{eq:scenario_OCP_dynamic_constraints_1}-\ref{eq:scenario_OCP_constraints}) are in place only for $ k  \in \mathbb{S}$, is $\{\bm{u}_{0:H}^\star, \overline{J_H}^\star\}$. In the following, we will refer to such a set $\mathbb{S}$ as a support sub-sample. In that case, the following result provides probabilistic guarantees for the maximum cost and the satisfaction of the constraints (\ref{eq:constraints}) when $\bm{u}_{0:H}^\star$ is applied to the unknown system.
	
	\begin{theorem}
		\label{theo:constraint_violation}
		Assume a support sub-sample $\mathbb{S}$ with cardinality $s$ is known. Let $V_c$ denote the probability that the incurred cost (\ref{eq:cost}) exceeds $\overline{J_H}^\star$ or that the constraints (\ref{eq:constraints}) are violated when the input trajectory $\bm{u}_{0:H}^\star$ is applied to the unknown system. For a given confidence parameter $\beta \in (0,1)$, under Assumptions \ref{as:sample_from_prior}, \ref{as:iid_posterior} and \ref{as:solver} it holds that
		\begin{equation}
			\mathbb{P}\left(V_c \leq 1-\epsilon(s)  \right)> 1-\beta,
		\end{equation}
		where $\epsilon(s)$ is the unique solution over the interval $(0,1)$ of the polynomial equation in the $v$ variable
		\begin{equation}\label{eq:polynomial_equation}
			\binom{K}{s}(1-v)^{K-s}-\frac{\beta}{K}\sum_{m=s}^{K-1}\binom{m}{s}(1-v)^{m-s}=0.
		\end{equation}
		
		\proof
		Due to Assumptions \ref{as:sample_from_prior} and \ref{as:iid_posterior}, the scenarios $\bm{\delta}^{[1:K]}$ generated by Algorithm~\ref{alg:scenario_generation} are independent samples from the respective distributions. Due to Assumption \ref{as:solver} and after combining (\ref{eq:scenario_OCP_dynamic_constraints_1}-\ref{eq:scenario_OCP_constraints}) to a single scalar-valued constraint, \cite[Theorem 1]{garatti2021} is applicable. The result follows since, for a fixed $K$, the implied function $\epsilon(k)$ monotonically decreases for an increasing $k$.
		\endproof
	\end{theorem}
	
	Thus, we need to find a support sub-sample $\mathbb{S}$ to obtain formal guarantees for the input trajectory $\bm{u}_{0:H}^\star$. The strength of the guarantees depends on the number of scenarios $K$ and the cardinality $s$ of the support sub-sample. The guarantees become stronger for a fixed $K$ as $s$ decreases. However, determining the support sub-sample with the smallest possible cardinality and, thus, the strongest possible guarantees is a complex combinatorial problem. Therefore, heuristics, e.g., the greedy algorithm presented in \cite{campi2018}, are usually employed to determine a possibly suboptimal $\mathbb{S}$. For a fixed cardinality $s$, increasing the number of scenarios $K$ leads to stronger guarantees. Therefore, additional scenarios are usually generated if the robustness level is insufficient for the intended application. However, since the problem (\ref{eq:scenario_OCP}) is generally non-convex due to the nonlinear dynamics, and it is thus not guaranteed that additional scenarios will not also increase $s$, it is impossible to determine the number of scenarios required for a given robustness level a priori.
	
	With Theorem \ref{theo:constraint_violation} and a suitable strategy to find a support sub-sample $\mathbb{S}$, it is possible to provide formal performance and constraint satisfaction guarantees. Due to the dedicated scenario generation process, these guarantee robustness to aleatory and epistemic uncertainties. The scenario-based formulation thus prevents the solution of the OCP from becoming overconfident in the face of uncertainty, allowing the use of the proposed method for safety-critical applications.
	
	\section{Simulation}
	\label{sec:simulation}
	In this section, we demonstrate the effectiveness of the proposed optimal control approach in a simulation\footnote{The code is available at \url{https://github.com/TUM-ITR/PGopt}}. The simulation setup is outlined in Sec.~\ref{sub:setup}. The proposed control approach is evaluated in Sec.~\ref{sub:basis_function}, assuming a known parametric model for the unknown system. In Sec.~\ref{sub:GP_approximation}, we demonstrate that it performs well even if no parametric model for the state transition function is known, using methods from Gaussian process regression.
	
	\subsection{Setup}
	\label{sub:setup}
	Consider a system of the form (\ref{eq:system}) with the unknown state transition function 
	\begin{equation}
		\bm{f}\left(\bm{x},u\right) = \begin{bmatrix}
			0.8x_1-0.5x_2+0.1\cos(3x_1)x_2\\
			0.4x_1+0.5x_2+(1+0.3\sin(2x_2))u
		\end{bmatrix} 
	\end{equation}
	and unknown process noise distribution
	\begin{equation}
		\bm{v}_t \sim \mathcal{N} \left( \bm{0}, \begin{bmatrix}
			0.03 & -0.004\\
			-0.004 & 0.01
		\end{bmatrix} \right).
	\end{equation}
	In order to simplify the interpretation of the simulation results, we assume that the observation function $g(\bm{x},u)=x_1$ and the distribution of the measurement noise $w_t \sim \mathcal{N}(0,0.1)$ are known. Note that this assumption is without loss of generality since an unknown observation model can be absorbed into the state transition model when the number of states is increased to $n_x+n_y$ \cite[Sec.~3.2.1]{frigola2015}.
	
	For the evaluation of the proposed optimal control approach, we consider the cost function $J_H = \sum_{t=0}^{H}u_t^2$ over the horizon $H=40$. In order to investigate how the control approach reacts to temporarily active constraints, we consider the constraints $2 \leq y_{20:25} $ and $|u| \leq 5$. For the scenario generation, $|T|=2000$ input-output measurements are obtained by simulating the system forward with a random input trajectory $u_t \sim \mathcal{N}(0,3 )$ starting from a random initial state with known distribution $\bm{x}_{T} \sim \mathcal{N}([ 2,\: 2]^{\mathsf T}, \bm{I}_2)$. 
	
	We employ the approach presented in \cite{svensson2017} to infer the model parameters and the latent state trajectory. In this approach, it is assumed that the state transition function $\bm{f}(\cdot)$ is a linear combination of $n_a$ basis functions $\bm{\varphi}(\bm{x}_t,\bm{u}_t)$ and that the process noise is normally distributed with zero mean. This yields the model $\bm{x}_{t+1} = \bm{A} \bm{\varphi}(\bm{x}_t,\bm{u}_t) +\bm{v}_t,\:\bm{v}_t \sim \mathcal{N}(\bm{0},\bm{Q})$ with unknown parameters $\bm{\theta} = \{\bm{A}\in \mathbb{R}^{n_x \times n_a},\bm{Q}\in \mathbb{R}^{n_x \times n_x}\}$. For $\bm{Q}$, an inverse Wishart (IW) prior with $\ell$ degrees of freedom and positive definite scale matrix $\bm{\Lambda}$ is assumed. The IW distribution is a distribution over real-valued, symmetric, positive definite matrices. For $\bm{A}$ a matrix normal (MN) prior with mean matrix $\bm{M} \in \mathbb{R}^{n_x \times n_a}=\bm{0}$, right covariance matrix $\bm{U} \in \mathbb{R}^{n_x \times n_x} = \bm{Q}$, and left covariance matrix $\bm{V} \in \mathbb{R}^{n_a \times n_a}$, is assumed. This conjugate prior results in a closed-form expression for $p(\bm{\theta} \mid \bm{x}_{T:-1}, \mathbb{D})$ \cite{svensson2017} and thus the efficient PG sampler presented in \cite{lindsten2012} can be used to draw samples from $p(\bm{\theta}, \bm{x}_{T:-1} \mid \mathbb{D})$. The selected parameters are given in Table~\ref{tab:parameters}.
	
	\begin{table}[t]
		\centering
		\begin{tabular}{c c c c c c}
			$\bm{\Lambda}$, $\bm{Q}^{\text{init}}$& $\ell$ & $\bm{A}^{\text{init}}$ & $K_b$ & $N$ \\
			\hline
			$100 \bm{I}_2$  & 10 & $\bm{0}$ & 1000 & 30
		\end{tabular}
		\caption{Parameters of the prior and the PG sampler.}
		\label{tab:parameters}
		\vspace*{-0.2cm}
	\end{table} 
	
	\subsection{Optimal control with known basis functions}
	\label{sub:basis_function}
	In order to show the applicability of the guarantees derived in this paper, we first assume that the basis functions $\bm{\varphi}(\bm{x},u)=[ x_1,\: x_2,\: u,\: \cos(3x_1)x_2,\: \sin(2x_2)u]^{\mathsf T}$ are known and that only the model parameters $\bm{\theta}=\{\bm{A}, \bm{Q}\}$ must be inferred. In order to facilitate the PG sampler's exploration of the posterior distribution, we scale the basis functions with the factors $[ 0.1,\: 0.1,\: 1,\: 0.01,\: 0.1]^{\mathsf T}$. For the prior for the corresponding weights, we choose $\bm{V} = 10 \bm{I}_5$. 
	
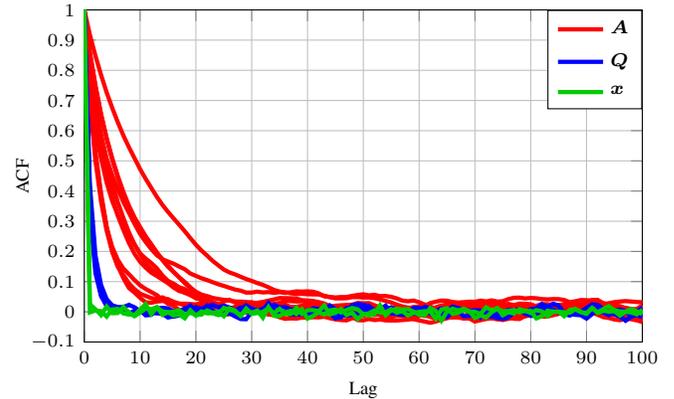
\begin{figure}[t]
		\pgfplotsset{width=9cm, compat = 1.18, 
			height = 6cm, grid= major, 
			legend cell align = left, ticklabel style = {font=\scriptsize},
			every axis label/.append style={font=\scriptsize},
			legend style = {font=\scriptsize},
		}
		\def\file{data/autocorrelation_seed_82.txt}
		
		\centering
		\begin{tikzpicture}
			\begin{axis}[
				grid=both,
				xmin=0, xmax=100,
				ymin=-0.1, ymax=1,
				xtick distance=10,
				ytick distance=0.1,
				ylabel=ACF, xlabel=Lag,
				set layers=standard,
				legend style={font=\scriptsize, at={(1,1)},anchor=north east, row sep=2pt},
				ylabel shift = -6 pt]
				
				\addplot[ultra thick,red] table[x=lag,y=A1]{\file};
				\addlegendentry{$\bm{A}$}
				\addplot[forget plot, ultra thick,red] table[x=lag,y=A2]{\file};
				\addplot[forget plot, ultra thick,red] table[x=lag,y=A3]{\file};
				\addplot[forget plot, ultra thick,red] table[x=lag,y=A4]{\file};
				\addplot[forget plot, ultra thick,red] table[x=lag,y=A5]{\file};
				\addplot[forget plot, ultra thick,red] table[x=lag,y=A6]{\file};
				\addplot[forget plot, ultra thick,red] table[x=lag,y=A7]{\file};
				\addplot[forget plot, ultra thick,red] table[x=lag,y=A8]{\file};
				\addplot[forget plot, ultra thick,red] table[x=lag,y=A9]{\file};
				\addplot[forget plot, ultra thick,red] table[x=lag,y=A10]{\file};
				
				\addplot[ultra thick,blue] table[x=lag,y=Q1]{\file};
				\addlegendentry{$\bm{Q}$}
				\addplot[forget plot, ultra thick,blue] table[x=lag,y=Q2]{\file};
				\addplot[forget plot, ultra thick,blue] table[x=lag,y=Q3]{\file};
				\addplot[forget plot, ultra thick,blue] table[x=lag,y=Q4]{\file};
				
				\addplot[ultra thick,black!20!green] table[x=lag,y=x1]{\file};
				\addlegendentry{$\bm{x}$}
				\addplot[forget plot, ultra thick,black!20!green] table[x=lag,y=x2]{\file};
				
			\end{axis}
		\end{tikzpicture}
		\vspace*{-0.8cm}
		
		\caption{Normalized auto-correlation function (ACF) between successive samples of the PG sampler without thinning. The red lines represent the ACF for the 10 different entries of the weight matrix $\bm{A}$, the blue lines represent the ACF for the 4 different entries of the process noise covariance matrix $\bm{Q}$, and the green lines represent the ACF for the 2 different entries of the state $\bm{x}_{-1}$.}
		\label{fig:autocorrelation}
	\end{figure}

	First, we ensure the validity of Assumption~\ref{as:iid_posterior}. For this purpose, we draw $K=10000$ samples from the posterior distribution over model parameters and latent state trajectories using a PG sampler without thinning. The normalized auto-correlation functions (ACFs) of consecutive samples of the model parameters $\bm{\theta}$ and the state $\bm{x}_{-1}$ are shown in Figure~\ref{fig:autocorrelation}. The ACF for all 16 parameters decays significantly for a lag of 50. Therefore, a thinning procedure with $k_d=50$ is used in the following to generate approximately independent samples.
	
	Afterward, $K=200$ scenarios are generated using Algorithm~\ref{alg:scenario_generation}. These scenarios are used to formulate an OCP as described in Sec.~\ref{sub:optimal_control}, which is then solved using the trajectory optimizer ALTRO \cite{howell2019}. In order to find a support sub-sample $\mathbb{S}$, required for the application of Theorem~\ref{theo:constraint_violation}, we first sort the scenarios based on their minimum distance to the constraint boundary. We then employ the greedy algorithm presented in \cite{campi2018}, which iterates over the scenarios (starting with the scenarios with the largest distance to the constraint boundary) and checks whether removing the constraints associated with a scenario changes the solution. The corresponding scenario is permanently removed from the constraint set if the solution remains unchanged. Based on the cardinality $s$ of the obtained support sub-sample, we compute the guarantees according to Theorem~\ref{theo:constraint_violation} with confidence $1-\beta = \SI{99}{\percent}$, whereby we use the algorithm presented in \cite{garatti2022} to solve the polynomial equation (\ref{eq:polynomial_equation}). Since, in this case, there is no uncertainty about the incurred cost $J_H$, as it does not depend on the uncertain dynamics, $1-\epsilon$ directly corresponds to an upper bound on the probability that the constraints are violated. After computing the guarantees, we apply the obtained trajectory $\bm{u}^\star_{0:H}$ to the actual system for validation. 
	
	In order to obtain meaningful results, the process of computing the optimal input trajectory and guarantees is repeated 100 times. Since our approach can also account for epistemic uncertainties, a new training dataset $\mathbb{D}$ is generated for each run as described in Sec.~\ref{sub:setup}. Of the 100 runs, in 19 cases, a locally optimal input trajectory $\bm{u}^\star_{0:H}$ is not found within the specified maximum number of iterations of the solver. Possible causes for the solver to fail are that the posed problem is infeasible due to the randomly generated initial state of the OCP or a high uncertainty in this run, or that the solver converges to an infeasible local minimum. The latter problem could be addressed by repeating the optimization with a different initialization. If the optimization problem is infeasible, a potential solution is to reduce the uncertainty by adding more training data or choosing a more specific prior. It may also help to reduce the number of scenarios and, therefore, the strength of the achievable guarantees. Only the 81 runs with a successful optimization are considered in the following.
	
	\begin{figure}[t]
		\pgfplotsset{width=9cm, compat = 1.18, 
			height = 6cm, grid= major, 
			legend cell align = left, ticklabel style = {font=\scriptsize},
			every axis label/.append style={font=\scriptsize},
			legend style = {font=\scriptsize},
		}
		\def\file{data/result_known_basis_functions_seed_82.txt}
		
		\centering
		\begin{tikzpicture}
			\begin{axis}[
				grid=none,
				xmin=0, xmax=40,
				ymin=-6, ymax=7.5,
				xtick={0,5,10,15,20,25,30,35,40},
				ytick={-6,-4,-2,0,2,4,6,8,10},
				ylabel=$y$, xlabel=$t$,
				set layers=standard,
				reverse legend,
				legend style={font=\scriptsize, at={(1,1)},anchor=north east, row sep=2pt},
				ylabel shift = -6 pt]
				
				\addplot[name path=A, forget plot, thick, opacity=0.2] table[x=t,y=y_opt_max]{\file};
				\addplot[name path=B, thick, opacity=0.2] table[x=t,y=y_opt_min]{\file};
				\tikzfillbetween[of=A and B]{opacity=0.2};
				\addlegendentry{$\{y_{0:H}^{[1:K]}\}$}
				
				\addplot[ultra thick,black!20!green] table[x=t,y=y_opt_mean]{\file};
				\addlegendentry{$\frac{1}{K}\sum\limits_{k=1}^K y_{0:H}^{[k]}$}
				
				\addplot[ultra thick, blue] table[x=t,y=y_sys]{\file};
				\addlegendentry{$y_{0:H}$}
				
				\draw [fill=red, fill opacity=0.2,red, opacity=0.2] (20,-6) rectangle (25,2); 
				
				\node[] at (axis cs: 22.5,-2) {\scriptsize{$h>0$}};
				
			\end{axis}
		\end{tikzpicture}
		\vspace*{-0.4cm}
		
		\caption{Example of the optimal control with known basis functions. The red area shows the output constraints, the gray area encompasses the 200 scenarios that were used to determine the input trajectory, the green line shows the mean prediction, and the blue line shows one realization of the output of the actual system when the input trajectory $\bm{u}^\star_{0:H}$ is applied from time $t=0$.}
		\label{fig:known_basis_functions}
	\end{figure}
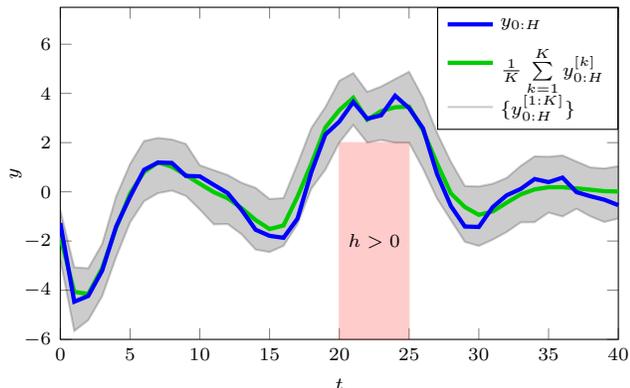
	
	Figure~\ref{fig:known_basis_functions} shows the results of an exemplary run. Initially, the control input is set to zero as this minimizes the stage cost, and the sampled dynamics converge to the stable equilibrium of the unforced system with output $y=0$. The system must then deviate from the equilibrium to satisfy the constraints $y_{20:25} \geq 2$. The input trajectory is chosen such that some of the scenarios precisely reach the constraint boundary and is thus not overly conservative. After the constraints are no longer present, the control input is again set to zero, and the sampled dynamics converge to the equilibrium again. The greedy algorithm yields a support sub-sample with cardinality $s=7$ in this run. According to Theorem~\ref{theo:constraint_violation}, with confidence $1-\beta = \SI{99}{\percent}$, the probability that the constraints are violated is thus less than $1-\epsilon = \SI{10.22}{\percent}$. The constraints are satisfied when the computed input trajectory is applied to the system.
	
	The results of all 81 runs with a successful optimization are summarized in Table~\ref{tab:results_known_BFs}. The average upper bound for the probability of constraint violation is $1-\epsilon=\SI{11.02}{\percent}$. In the forward simulation, the constraints are violated in \SI{4.94}{\percent} of the runs. Thus, the theoretical robustness levels presented in this paper appear slightly weaker than the actual robustness. Although the theoretical guarantees in this simulation may not be sufficient for every application, the ability to compute guarantees emphasizes the solid theoretical foundation of our approach. To the best of our knowledge, no other approaches allow the derivation of such guarantees under comparable assumptions. If stronger guarantees are desired, additional scenarios can be considered.
	
	The specified runtimes are obtained on an \SI{2.25}{\giga \hertz} 64-core AMD EPYC 7742 processor, on which the individual runs are executed CPU-parallelized, i.e., they correspond to a single core. As expected, the computational complexity of the approach is comparatively high, especially for the computation of the theoretical guarantees. This is partially due to the fact that solving chance-constrained optimization problems is, in general, computationally intensive. In addition, it should be noted that a very general class of nonlinear dynamical systems is considered and that the approach provides theoretical guarantees even with little prior knowledge and under mild assumptions.
	
	\begin{table}[t]
		\centering
		\begin{tabular}{|c| c|}
			\hline
			max. constraint violation probability $1-\epsilon$ & \SI{11.02}{\percent} $\pm $ \SI{2.53}{\percent}\\
			\hline
			number of runs with constraint violations & $4\ (=\SI{4.94}{\percent})$ \\
			\hline
			$J_H$ & $26.61\ \pm \ 10.93$\\
			\hline
			time to generate scenarios & \SI{900}{\second} $\pm $ \SI{56}{\second}\\
			\hline
			time to compute $\bm{u}^\star_{0:H}$\vspace{1pt} & \SI{366}{\second} $\pm $ \SI{257}{\second}\\
			\hline
			time to compute $1-\epsilon$ & \SI{15629}{\second} $\pm $ \SI{2118}{\second}\\
			\hline
		\end{tabular}
		\caption{Results (mean $\pm$ std) of the 81 successful runs with known basis functions.}
		\label{tab:results_known_BFs}
		\vspace*{-0.2cm}
	\end{table}
	
	\subsection{Optimal control with generic basis functions}
	\label{sub:GP_approximation}
	In the following, we show that the proposed optimal control approach can yield good results even if no parametric model is known, which might be the case in practice. Instead of the actual basis functions, we use the reduced-rank GP approximation proposed in \cite{solin2020} to systematically determine the basis functions $\bm{\varphi}(\bm{x},\bm{u})$ and the parameter $\bm{V}$ of the prior. We choose a GP with a squared exponential kernel and select the hyperparameters of the GP and the approximation based on the training data. These parameters are given in Table \ref{tab:GP_parameters}. Then, $K=100$ scenarios are generated using a PG sampler with the same parameters as in the previous example, and the resulting OCP is solved as before. Again, the procedure is repeated 100 times, of which the optimizer finds an optimal solution in 77 cases.
	
	\begin{table}[t]
		\begin{center}
			\begin{tabular}{c c c c c c}
				$l$ & $s_f$ & $m_{x_1}, m_{x_2}, m_{u}$ & $L_{x_1},L_{x_2}$ & $L_{u}$ \\
				\hline
				2 & 100 & 5 & 20 & 10
			\end{tabular}
			\caption{Parameters of the reduced-rank GP approximation.}
			\label{tab:GP_parameters}
		\end{center}
		\vspace*{-0.2cm}
	\end{table}
	
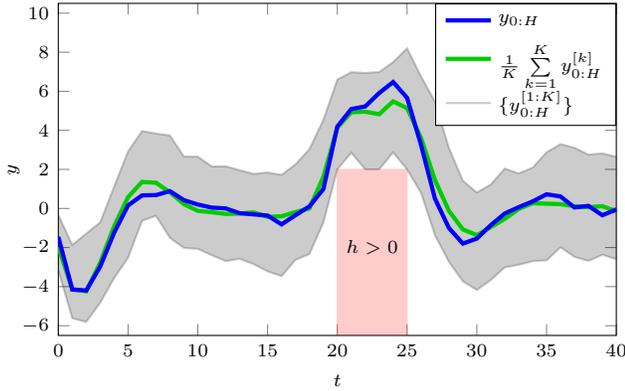
\begin{figure}[t]
		\pgfplotsset{width=9cm, compat = 1.18, 
			height = 6cm, grid= major, 
			legend cell align = left, ticklabel style = {font=\scriptsize},
			every axis label/.append style={font=\scriptsize},
			legend style = {font=\scriptsize},
		}
		\def\file{data/result_generic_basis_functions_seed_82.txt}
		
		\centering
		\begin{tikzpicture}
			\begin{axis}[
				grid=none,
				xmin=0, xmax=40,
				ymin=-6.5, ymax=10.5,
				xtick={0,5,10,15,20,25,30,35,40},
				ytick={-8,-6,-4,-2,0,2,4,6,8,10},
				ylabel=$y$, xlabel=$t$,
				set layers=standard,
				reverse legend,
				legend style={font=\scriptsize, at={(1,1)},anchor=north east, row sep=2pt},
				ylabel shift = -6 pt]
				
				\addplot[name path=A, forget plot, thick, opacity=0.2] table[x=t,y=y_opt_max]{\file};
				\addplot[name path=B, thick, opacity=0.2] table[x=t,y=y_opt_min]{\file};
				\tikzfillbetween[of=A and B]{opacity=0.2};
				\addlegendentry{$\{y_{0:H}^{[1:K]}\}$}
				
				\addplot[ultra thick,black!20!green] table[x=t,y=y_opt_mean]{\file};
				\addlegendentry{$\frac{1}{K}\sum\limits_{k=1}^K y_{0:H}^{[k]}$}
				
				\addplot[ultra thick, blue] table[x=t,y=y_sys]{\file};
				\addlegendentry{$y_{0:H}$}
				
				\draw [fill=red, fill opacity=0.2,red, opacity=0.2] (20,-8) rectangle (25,2); 
				
				\node[] at (axis cs: 22.5,-2) {\scriptsize{$h>0$}};
				
			\end{axis}
		\end{tikzpicture}
		\vspace*{-0.4cm}
		
		\caption{Example of the optimal control with generic basis functions. The red area shows the output constraints, the gray area encompasses the 100 scenarios that were used to determine the input trajectory, the green line shows the mean prediction, and the blue line shows one realization of the output of the actual system when the input trajectory $\bm{u}^\star_{0:H}$ is applied from time $t=0$.}
		\label{fig:GP_approximation}
	\end{figure}

	Figure~\ref{fig:GP_approximation} shows the results of an exemplary run with generic basis functions. The resulting trajectories are similar to the case with known basis functions. Only the uncertainty intervals are larger, which is expected as less specific prior knowledge is brought in. Due to the higher uncertainty, a more conservative input trajectory is chosen, and the distance of the actual output signal to the constraint boundary is larger.
	
	The results of all 77 successful runs with generic basis functions are summarized in Table~\ref{tab:results_generic_BFs}. In this case, no formal guarantees for the constraint satisfaction can be derived since Assumption~\ref{as:sample_from_prior} is not satisfied as the employed basis functions cannot represent the actual dynamics with arbitrary precision. Nevertheless, the resulting input trajectories are robust, and when applied to the actual system, the constraints are violated only in two runs. The higher cost compared to the case with known basis functions is due to the fact that a more conservative input trajectory is chosen due to the higher uncertainty.
	
	\begin{table}[t]
		\centering
		\begin{tabular}{|c| c|}
			\hline
			number of runs with constraint violations & $2\ (=\SI{2.60}{\percent})$ \\
			\hline
			$J_H$ & $81.81\ \pm \ 292.5$\\
			\hline
			time to generate scenarios & \SI{14138}{\second} $\pm $ \SI{200}{\second}\\
			\hline
			time to compute $\bm{u}^\star_{0:H}$\vspace{1pt} & \SI{749}{\second} $\pm $ \SI{1108}{\second}\\
			\hline
		\end{tabular}
		\caption{Results (mean $\pm$ std) of the 77 successful runs with generic basis functions.}
		\label{tab:results_generic_BFs}
		\vspace*{-0.2cm}
	\end{table} 
	
	As shown in this subsection, the proposed approach is thus able to provide well-functioning and robust input trajectories based on a prior that can be formulated using the well-known GP framework, even without knowledge of a parametric model of the dynamics.
	
	\section{Conclusion}
	\label{sec:conclusion}
	This paper presents a novel approach for the optimal control of unknown dynamical systems with latent states. Based on a prior for the dynamics and the latent state trajectory, samples from the posterior distribution over future trajectories of the unknown system are drawn using PMCMC methods. These samples are then used to formulate an optimal control problem that can be solved using well-known methods. Probabilistic performance and constraint satisfaction guarantees are derived for the obtained input trajectory by employing a scenario perspective.

	\bibliographystyle{IEEEtran}
	\bibliography{ms}
	
\end{document}